\documentstyle[12pt,aasms]{article}
\def\mincir{\raise -2.truept\hbox{\rlap{\hbox{$\sim$}}\raise5.truept
\hbox{$<$}\ }}

\def\magcir{\raise -2.truept\hbox{\rlap{\hbox{$\sim$}}\raise5.truept
\hbox{$>$}\ }}
\def\minmag{\raise-2.truept\hbox{\rlap{\hbox{$<$}}\raise 6.truept\hbox
{$>$}\ }}

  \def\Mpch{\, h^{-1}{\rm Mpc}}    

\begin{document}

\title{VOID ANALYSIS AS A TEST FOR DARK MATTER COMPOSITION? }

\author{Sebastiano Ghigna\altaffilmark{1} \& Silvio A. Bonometto}
\affil{Dipartimento di Fisica dell'Universit\`a di Milano,
Via Celoria 16, I-20133 Milano, Italy and \\
INFN, Sezione di Milano, Italy}

\author{J\"org Retzlaff \& Stefan Gottl\"ober}
\affil{Astrophysikalisches Institut Potsdam, An der Sternwarte 16, 
D--14482 Potsdam, Germany}

\author{Giuseppe Murante}
\affil{Istituto di Cosmogeofisica del C.N.R.,
C.so Fiume 4, 10133 Torino, Italy}
   
\altaffiltext{1}{Physics Department, University of Durham, 
Science Laboratories, South Road, Durham DH1 3LE, England (present address)}

\begin{abstract}

We use the void probability function (VPF) to compare the redshift--space
galaxy distribution in the Perseus--Pisces redshift survey with 
artificial samples extracted from $N$--body simulations of standard cold
dark matter (CDM) and broken scale invariance (BSI)
models. Observational biases of the real data set are reproduced as well as
possible in the simulated samples. Galaxies are identified as residing in peaks
of the evolved density field and overmerged structures are fragmented into
individual galaxies in such a way as to reproduce both the correct luminosity
function and the two--point correlation function (assuming suitable $M/L$
values). Using a similar approach, it was recently shown that the VPF can
  discriminate between CDM and a cold$+$hot dark matter (CHDM) model
  with $\Omega_{{\rm cold}}/\Omega_{{\rm hot}}/\Omega_{{\rm baryon}}=0.6/0.3/0.1$.
Our main result is that both CDM (as
expected from a previous analysis) and BSI fit observational data. The
robustness of the result is checked against changing the observer's position in
the simulations and the galaxy identification in the evolved density field.
Therefore, while the void statistics is sensitive to the passage from CDM to
CHDM (different spectrum and different nature of dark matter), it is not to 
the passage from CDM to BSI (different spectrum but same dark matter). On such 
a basis, we conjecture that the distribution of voids could be directly 
sensitive to the
nature of dark matter, but scarcely sensitive to the shape of the transfer function. 
\end{abstract}

\keywords{cosmology: theory -- dark matter -- 
galaxies: clustering -- large--scale
structure of the Universe }
 
\section{Introduction}

The Void Probability Function (VPF) is used as a
statistical tool to explore the properties of the Large Scale Structure (LSS)
of galaxies. This statistics provides a quantitative estimate of the
probability of finding empty regions in the galaxy distribution and gives
information on the LSS that cannot be predicted from the observed low order
correlation functions (White 1979) and that is however beyond the content of
measurable correlation functions for any finite part of the Universe. The void
statistics has been analyzed for many galaxy samples: the Southern Sky
redshift survey (Maurogordato, Schaeffer \& da Costa 1992), the 1.2 Jy 
IRAS redshift survey
(Bouchet et al. 1993), the Center for Astrophysics survey (CfA; 
Vogeley et al. 1992 and 1994). 
A preliminary version
of the Perseus--Pisces Survey (PPS; Giovanelli \& Haynes 1991, and references
therein) was studied  by Fry et al. (1989), who also made a comparison
with CDM $N$--body simulations. Other works on the VPF for observational
data include examination of the sky--projected galaxy distribution (Sharp 1981;
Bouchet \& Lachi\`eze--Rey 1986) and the distribution of clusters of
galaxies (Huchra et al. 1990; Jing 1990; Cappi, Maurogordato \& 
Lachi\`eze--Rey 1991). Theoretical properties of the VPF in the
framework of the hierarchical scaling model (HS) have been considered by Fry
(1986). A VPF analysis of $N$--body simulations has been carried out by 
Einasto et al. (1991) and Weinberg \& Cole (1992), who used the VPF to
discriminate between Gaussian and non--Gaussian initial conditions. Little
\& Weinberg (1994) investigated the effects of
varying the biasing prescription used to identify ``galaxies'' in the
simulations. Vogeley et al. (1994) compared the VPF for the CfA and for 
various $N$--body simulations, including three variants of the CDM model.
The void statistics of the CDM and CHDM models in the non--linear
  clustering regime was addressed by Bonometto et al. (1995), using 
PM simulations in a box of $50\Mpch$ with a cell side of
about $98\,h^{-1}$kpc, 
and confronted with HS predictions. 
The above simulations are described in detail by 
Klypin, Nolthenius \& Primack (1995) and
  were also considered by Ghigna et al. (1994; Paper A in the following), who  
estimated the VPF for artificial galaxy samples extracted from 
them and a sample of galaxies from the PPS database.

In Paper~A the close comparison between real and artificial samples showed that
the VPF discriminates between CDM and CHDM (with density parameters
$\Omega_{{\rm baryon}}/\Omega_{{\rm cold}}/\Omega_{{\rm hot}}= 
0.1/0.6/0.3$). 
It was also shown that the VPF is scarcely affected by the bias level
of CDM (i.e. by the {\sl amplitude} of its linear spectrum of density
fluctuations) and it was then suggested that the void statistics could be
chiefly determined by the composition of dark matter (DM) and/or
the shape of the spectrum. Here we 
want to explore its dependence on the latter point
further, by performing a similar analysis of the BSI and 
standard 
  CDM models, which differ in their power spectrum
but have the same DM composition.

The standard CDM model relies on the assumption of an 
Einstein-deSitter
universe ($\Omega_{{\rm baryon}} + \Omega_{{\rm cold}} = 1$) with $H_0 = 50$
km$\,$s$^{-1}$Mpc$^{-1}$
and a Harrison-Zeldovich spectrum of adiabatic
perturbations of the primordial density field. 
Normalizing the standard CDM model with the COBE data, it became
evident that this model is in conflict with observational data on
scales less than 10 Mpc (see e.g. Ostriker 1993). In models with broken
scale invariance of the initial density fluctuation spectrum this
spectrum is of Harrison-Zeldovich type only in the limits of small and
large scales, however on intermediate scales it is tilted. Such
spectra arise naturally in double inflationary scenarios (Gottl\"ober,
M\"uller, Starobinsky 1991). As was discussed by Gottl\"ober,
M\"ucket,  Starobinsky (1994), the parameters of the underlying
inflationary model can be chosen in such a way that the predictions of
the model are in agreement  with observational results. As in standard CDM 
two matter components ($\Omega_{{\rm baryon}} = 0.05$, 
$\Omega_{{\rm cold}} = 0.95$) and a Hubble constant $H_0 = 50$
  km$\,$s$^{-1}$Mpc$^{-1}$ are assumed in the BSI model.
 
A crucial point to test structure formation models by means of VPF in
dissipationless simulations is the galaxy identification scheme: changing the
efficiency of galaxy formation in low density areas bears an obvious impact on
the resulting VPF (Betancort--Rijo 1990; Einasto et al. 1991; Little \&
Weinberg 1994). As was discussed by Little \& Weinberg (1994), three criteria 
have been generally used to identify galaxies in the density field: 
(i) They can be set on the peaks of
the linear density field (linear biasing; e.g., Davis et al. 1985). (ii) They
can be set in high--density regions of the evolved density field. (iii) The
biasing relation derived by Cen \& Ostriker (1993) from their CDM hydrodynamic
simulations can be used. However, it is not clear whether the linear biasing
approach  yields the seeds where non--linear structures later form (Kates, 
Kotok \& Klypin 1991; Katz, Quinn, \& Gelb 1993), while the physical biasing 
relation of Cen \& Ostriker (1993) is derived only from CDM simulations 
spanning a limited dynamical range. 
Referring to the evolved density field seems then the most reliable 
prescription. Accordingly, as in Paper~A, we shall identify galaxies as 
corresponding to high--density peaks (our {\sl dark haloes} in the following) 
in the simulation volumes (criterion $(ii)\,$). More complicated variants of 
this prescription, however, have also been considered: e.g. one can
attempt to
estimate the thermal history of gas particles (even in the absence of
hydrodynamics, by following DM trajectories) and use {\sl cooled} particles
as tracers of galaxies
(Kates, Kotok \& Klypin 1991; Klypin \& Kates 1991; Kates et al. 1995). 

To properly address the subject of galaxy identification, we still 
have to face an intrinsic limitation of dissipationless simulations: these ones
are known to yield large haloes in central parts of
groups or clusters, with masses $M > 10^{13}M_{\odot}$, well beyond the galaxy
mass range (e.g., Gelb \& Bertschinger 1994). This {\sl overmerging}
 is partly due to lack of numerical resolution, but is also an effect of 
neglecting the 
dissipative processes which act on
galaxy mass scales. In the absence of 
these processes, tidal forces disrupt galaxy--size objects (e.g. 
  Moore, Katz \& Lake 1996).
In the real world, dissipation allows baryons to cool down and
  condense earlier forming potential wells of galactic size, before     
galaxies
are assembled in clusters. In this way substructures arise in time to prevent
overmerging. The recipe defined by criterion $(ii)$ needs then to be 
supplemented 
with further prescriptions, since many patterns can be
followed to assign galaxies to peaks, according to different ways to
fragment {\sl overmerged} structures into individual objects. A first general
requirement is that the galaxy identification scheme must agree with the
luminosity function and the two--point correlation function. 

In Paper~A, a method was devised to fragment dark haloes into galaxies, with
luminosities distributed according to a Schechter function (Schechter 1976), 
whose output depends
 on two parameters: (i) the average separation of the galaxy
population to be reproduced and (ii) the mass--to--light ratio $M/L$, 
assumed to be constant for all dark haloes in the simulation (whose
connection with the physical $M/L$ will be discussed shortly). 
This same method will be adopted for the present analysis. Let us outline that
a more detailed procedure could not be efficiently tested on data, as this
two--parameter fragmentation prescription already allows us to reproduce the correct
slope and amplitude for the galaxy two--point correlation function. 

Once galaxies are identified from simulations, their distribution in 
the computational volume is to be dealt
with in order to reproduce a {\sl sample } with characteristics similar to the
one extracted from PPS. The starting point is then to reduce the distribution
to an artificial
redshift--space galaxy set. It must have the same galaxy number density
as the real sample and must reproduce its geometrical shape, to account for
boundary effects. The reduction procedure 
depends on 
the choice of an
observer's viewpoint and {\sl observing} each simulation from several
viewpoints allows an estimate of the {\sl sky variance} within a given
real--space volume. The fit of simulations to real data will then be made
by considering a 
large set of {\sl observers}, in order 
to verify whether, within the estimated {\sl sky variance}, 
a galaxy sample with the same properties as the PPS one 
could be {\sl observed} in a world arising from the cosmological model 
considered.

This article is organized as follows. In Section 2, we present the VPF
statistics. In Section 3, we briefly review the
theoretical background of the BSI model and provide details on the simulations.
In Section 4, we describe the observational material and give the
characteristics of the galaxy sample we use for our analysis. In Section 5, the
procedure to reduce the simulations and create the artificial galaxy samples to
be compared with PPS is debated. The VPF analysis is performed
and its results are given in Section 6, while Section 7 is devoted to the
conclusions we draw from our analysis.

\section{The void probability function}

The VPF is a tool to characterize a spatial distribution of objects and
is defined as the probability, $P_0$, of finding no objects within a given 
randomly placed sampling volume $V_r$ (characterized by the scale $r$). The VPF conveys information about
correlations of any order. It can be shown (White 1979) that the
following relation holds:
\begin{equation}
P_0(r)=\exp\,\left[\sum_{q=1}^\infty {(-{\bar N_r})^q\over q!}\,\bar 
\xi_q (r)\right]
\label{eq:p0}
\end{equation}
where ${\bar N_r}$ is the average number of objects within $V_r$,
  $\bar \xi_1 (r) \equiv 1$ and $\bar \xi_q (r)$   are $q$--th order
correlation functions averaged over $V_r$ (in particular, ${\bar\xi_2}$
is the variance of counts, simply {\sl variance} in the following).
However, since $P_0$ depends only on the
  number of non--empty cells, regardless of      the number of objects
  contained inside them, it can be said  from a qualitative point of view 
that the VPF of a point distribution is related to its
geometry, rather than to its clustering.
For a completely uncorrelated
(i.e. Poissonian) distribution, it is $P_0(r)=\exp{(-\bar N_r)}$, so that 
any departure from the latter quantity
represents the signature for the presence of clustering.
Here, as in Paper~A, we use spheres of different radii $r$ to estimate
the VPF for (real and artificial) galaxy samples.  Spheres are completely
contained within the samples, i.e. their centers are positioned
at distance $\ge r$ from the sample boundaries. If $V_{\rm VLS}$ is the
volume of the sample, we take $N_r=2\,V_{{\rm
VLS}}/V_r$  spheres ($V_r=4\pi r^3/3$), where the factor of 2
accounts for the presence of clustering (Fry \& Gazta\~naga 1994).
We estimate sampling errors through the bootstrap method (e.g., Ling,
Frenk \& Barrow 1986; Efron \& Tibshirani 1991).
 
\section{Characteristics of BSI model and simulations}

The cold dark matter BSI model arises from a double-inflationary
scenario.  Two subsequent inflationary stages which are driven by a
$R^2$ term and a massive scalar field in the Lagrangian density lead
to a power spectrum of potential fluctuations which exhibits broken
scale invariance (cf.\ Gottl\"ober, M\"uller, \& Starobinski 1991).
Compared to a flat Harrison-Zeldovich spectrum resulting from 
single inflation, the BSI spectrum is characterized by a step
between the two asymptotically flat regions $k\rightarrow 0$ and
$k\rightarrow \infty$.  Besides the step's location at $k_{{\rm break}}$,
its relative height $\Delta$ constitutes the second free parameter in
addition to those of the standard CDM model.  From the inflationary scenario we
expect fluctuations with initially Gaussian statistics.  The transfer
function linearly maps the initial power spectrum of perturbations to
the present epoch. We use the transfer function for 
adiabatic fluctuations in a 
  CDM model with $\Omega_{\rm{tot}}=1$, $\Omega_{\rm baryon}=0.05$, and a
present Hubble constant $H_0=50\,{\rm{km\,s^{-1}\,Mpc^{-1}}}$,
as computed by Bond \& Efstathiou (1984).

Gottl\"ober, M\"ucket \& Starobinski (1994) compared a number of linear
predictions of the BSI model with the corresponding observational quantities.
In particular, they analyzed the variance of counts in cells of IRAS galaxies,
the two--point angular correlation function for the APM sample, bulk 
flow velocities, the Mach number,
 and studied the compatibility of the model with observed high--redshift
 galaxy and quasar number densities. From such comparisons they worked
out the following best fit values for the BSI parameters: 
$k_{\rm{break}}^{-1}=3\,{\rm{Mpc}}$ (for $h=0.5$ as in this paper) and
$\Delta=3$. Here we shall consider BSI simulations
run by Kates et al. (1995),
starting from a BSI spectrum with $k_{\rm{break}}^{-1}$
  and $\Delta$ as above normalized to the first--year COBE quadrupole data 
(Smoot et al. 1992), yielding a bias parameter 
$b_8\equiv\sigma^{-1}(8h^{-1} {\rm{Mpc}})=2.2$. 
  (A reanalysis of the simulation based on second--year COBE   data (G\'orski
et al. 1994) yields $b_8 = 1.7$).

Kates et al. (1995) already submitted such simulation outputs to
various statistical tests; in particular they found good fits for counts in
cells, the probability distribution function of matter, the mass function of
galaxies and clusters, their two-point spatial correlation function and
integral bias, streaming motions, and peculiar velocity dispersions. Amendola
et al. (1995) also found good fits of such simulations to the power
spectrum reconstructed from the CfA survey, as well as
to the angular correlation functions of APM galaxies and higher order moments 
of cell counts from APM-Stromlo data.

\section{Observational catalog}
 
  Observational data are provided by the Perseus--Pisces redshift survey
(for details see Giovanelli \& Haynes 1989 and 1991). The sample
we consider is limited to the region bound by $22^h \le 
\alpha \le 3^h\; 10^m$ and $0^\circ \le \delta \le 42^\circ\; 30^\prime$, in
order to exclude areas of sky of high galactic extinction.
Zwicky magnitudes ($m_{\rm{Zw}}$) are corrected for extinction using 
  the absorption maps of Burstein \& Heiles (1978), and all galaxies brighter than 
$m_{\rm{Zw}}=15.5$ are included.  The resulting sample is virtually 100\% 
complete for all morphological types to this limiting magnitude, and
contains 3395 galaxies.
The galaxy distribution described by these data is a pure redshift--space
one. The only correction applied to 
observed velocities consists in
subtracting our motion with respect to the Cosmic Microwave Background
radiation (CMB).
By this transformation the observer is put at rest in the frame of reference 
where the CMB dipole vanishes.

Our analyses are made on a volume--limited subsample (VLS) of the PPS survey,
with the limiting magnitude $M_{\rm lim} = -19+5\log h$, corresponding to $79\,
h^{-1}$Mpc for the limiting depth $d_{\rm lim}$. With the angular
boundaries given above, the geometry of the sample is that of a broad
wedge extending over $107.5^\circ$ in ascension and $42.5^\circ$ in
declination and therefore covering a volume 
$V_{{\rm VLS}} = 1.5\times 10^5 h^{-3}$Mpc$^3$. 
 The sample contains 902 galaxies with mean
galaxy separation $d_{\rm gal}=5.5\,h^{-1}$ Mpc. (This  sample differs from the one used
in Paper~A, which was obtained by correcting for our motion with respect to
the centroid of the Local Group and included 1032 galaxies with an average
separation of $5.2\,h^{-1}$ Mpc. An analysis of the CDM and CHDM simulations of
Paper~A compared to the present observational sample confirms the
conclusions of Paper~A and will be presented in a
forthcoming paper, Ghigna et al. 1996).

\section{Simulations and artificial samples}
 
We consider the BSI model with the parameters discussed in Section 3 
and, for comparison, an unbiased CDM model ($\sigma_{{\rm DM}}=1$ on the 8$\, 
h^{-1}$Mpc
scale; this normalization is within one standard deviation from the
amplitude detected by COBE on the quadrupole scale during its
  first  year of activity, and is compatible with the
second--year data). 
Simulations are performed with a Particle--Mesh code (PM)
for $75\, h^{-1}$ Mpc boxes with 
$128^3$ particles on a 256$^3$--cell  grid. This is well suited for
comparison with the observational VLS of limiting depth $d_{{\rm lim}} =
79\Mpch$. The cell size is $l_{\rm c} \simeq 0.29\Mpch$ corresponding
to an expected spatial resolution $\sim 3l_{\rm c} \simeq 0.88\Mpch$.
We took our simulations from those performed
by Kates et al. (1995).  Initial conditions were set at redshift $z =
25$, using the CDM transfer function from Bond \& Efstathiou
  (1984). The COBE--normalized perturbation spectra are shown in Figure 1.

The simulations provide the positions of the particles at the present epoch.
We employ the TSC interpolation scheme (Hockney \& Eastwood 1981)
to obtain the density field on the grid.
In this set of data, we identify cells which correspond to local
density
maxima (peaks) exceeding a suitable overdensity threshold $w_{\rm th}$, whose
 value depends on the model and on the galaxy identification scheme
used (through the $M/L$ parameter). 
  Each selected peak, given the mass of the
$3^3$ surrounding cells, is a dark matter {\sl halo} 
and is assigned a {\sl radius }$R_h \simeq 0.55 \Mpch$ ($R_h$ is defined
such that $4\pi R_h^3/3 \equiv 27l_{\rm c}^3$; 
here it exceeds the corresponding quantity of Paper~A
by a factor of 3, the difference
being due to the joint effects of the reduced spatial resolution, a
factor of 2, and the larger box size, a factor of 1.5). 
Any such halo is not a single {\sl galactic halo}.
According to its size, it is 
expected to be a group, housing a certain number of
galaxies (each one carrying a {\sl galactic halo} of its own). 
Small haloes can also be expected to yield only one galaxy, with its 
{\sl galactic halo} located somewhere inside the $R_h$ sphere we do not 
resolve. 

Once haloes are singled out in the simulation box, our purpose is then to
  assign  a population of galaxies to   them  analogous to that contained
in the observational sample from PPS. Henceforth, from haloes we
aim to obtain galaxies brighter than $L_{{\rm lim}} = 3.06 \times
10^9h^{-2}L_\odot$ (corresponding to
$M_{\rm lim}$ of Section 4) with average separation $d_{\rm gal}$. To do this, 
we first calculate the expected total luminosity
$L_{\rm total}$ of galaxies in the computational volume $V=l_{\rm box}^3=
(75\Mpch)^3$.
Let us take a Schechter luminosity function (Schechter 1976) 
$\phi(L)\,dL=\phi_*\,({L /L_*})^\alpha\,\exp (-{L / L_*})\,d(L/L_*)$, 
  with $\alpha=-1.07$, and $L_*$ being the
luminosity of galaxies whose absolute magnitude is $M_*=-19.68$ (Efstathiou,
Ellis \& Peterson 1988). Moreover, 
we take $\phi_*=1.17\times 10^{-2} h^{3} $Mpc$^{-3}$,
so as to obtain the correct galaxy separation $d\,$ from 
the {\sl normalization condition} $\int_{L_{{\rm lim}}}^\infty 
  \phi(L) \,dL \equiv d_{\rm gal}^{-  3}$. The expected total luminosity is then 
given by the integral ${L}_{\rm t}=l^3 \int_{L_{\rm{lim}}}^\infty L\,\phi(L) 
\,dL$. For our box, $L_{{\rm t}} = 1.55\times 10^{13} h^{-2} L_\odot$. 

The next step amounts to assuming a suitable mass--to--light ratio $M/L$
(which is the effective parameter in our fitting procedure) for our galaxy
population and calculating the total mass expected in the box $M_{\rm
t}=(M/L)\times L_{\rm t}$. Then, for a fixed $M/L$, we select the most massive
$N_{\rm hal}$ haloes, so that $\sum_{k=1}^{N_{\rm hal }}M_k = M_{\rm t}$ (in
order to facilitate galaxy allocation in peaks as described below, we
give the peaks a mass exceeding $M_{\rm t}$ by $\sim 5$\%; this is needed
because of the finite luminosity of the faintest galaxies considered and 
yields slight variations, peak by peak, of the effective $M/L$, whose 
relative standard deviation is $\sim 1.5\, \%$).
Afterwards, we produce a realization of the mass
function $n(M)\,dM = \phi(L)\,dL$ with $M$ and $L$ related through the 
$M/L$ ratio we have fixed.  
This amounts to generating a set of values for
the masses of $N_{\rm gal}=(l_{\rm box}/d_{\rm gal})^3$ 
galaxies. Again, if $M_i$ is the mass
assigned to the $i$--th galaxy, the following condition holds: 
$\sum_{i=1}^{N_{\rm gal}}M_i = M_{\rm t}$. Finally, such sets of ``galaxies''
are distributed among the DM haloes selected. We take the most massive galaxy
and assign it to the most massive halo in the simulation. Because the halo is
more massive than the galaxy, there is some halo mass left for another galaxy.
If the mass left in the halo is larger than the mass of the second largest
  galaxy, we {\sl assign} that galaxy to the halo. If not, then the next most
  massive galaxy is tried and so on until we find a galaxy  whose mass          
smaller than the remaining mass of the halo. 
This procedure is repeated until the mass of the most massive halo is
subdivided into galaxies. Then
we take the most massive galaxy left
and assign it to the second most massive halo, and operate on it following
the same steps as above. We end our procedure when all galaxies have been 
given a ``parent'' halo.

In this way, several haloes contain more than one galaxy. In the real 
world these
galaxies would have different redshifts because of their velocities inside the
halo to which they belong. This feature can be suitably reproduced by 
giving each
galaxy a velocity ${\bf v}_g = {\bf v}_i + {\bf \Delta v}_g$, where ${\bf v}_i$
is the global velocity of the $i$--th halo and ${\bf \Delta v}_g$ results from
local motions. We shall assume that local motions are approximately virialized.
Henceforth ${\bf \Delta v}_g$ shall have Gaussian--distributed components,
with variance $\langle {\bf \Delta v}_i^2 \rangle / 3$, where $\langle {\bf
\Delta v}_i^2 \rangle = G M_i/R_{h}\,$ ($M_{i}$ is the halo mass). This 
amounts to assume virial equilibrium within the halo radius $R_{h}$. 
We verified that such velocity corrections do not modify the small--scale
profile of the pairwise galaxy velocity dispersion in a significant way. 

It is important to outline that the meaning of the $M/L$
parameter is to be treated with much caution. The cell size
in the simulation has a critical impact on the value of $M/L$.
The numerical procedure carried out in the PM method has a smoothing
effect on forces, which become simply absent below a scale of
the order of $l_{\rm c}$, while in the real world 
there is a length scale $l_{\rm g}$, corresponding to a typical galactic
  mass  scale, below which the dynamics is dominated by dissipative forces. 
The values of $M/L$ worked out by fitting a Schechter function (Schechter 1976) to a
simulation and in the real world can be expected to be comparable 
only if $l_{\rm c}$ and $l_{\rm g}$ are of the same order of
magnitude.  For $l_{\rm c} \gg l_g$, 
as is the case
here, forces are smoothed over too large a scale: although the matter
distribution can be faithfully reproduced over scales greater
than $3l_{\rm c}$, 
the density contrast reached is never high enough to permit us
to give a direct physical significance to the values of $M/L$ that we shall
be working out. In spite of that, although the individual values of $M/L$
do not make sense, the ratios among values obtained for different
cosmological models do. As we shall show, standard CDM and BSI
yield different values of $M/L$ and from 
their ratio we can gain information on the physical $M/L$.

Once galaxies are defined and ${\bf v}_g$ is assigned to each of them, we
construct the galaxy distribution in     redshift space, for a given observer's
location. The depth of the VLS (whose geometry is described in Section 4)
slightly exceeds the size of the simulation (79 to 75),
but this difficulty can be easily overcome by having the axis of the
simulated VLS 
close to the direction of the box diagonal, which stretches
over 130$\, h^{-1}$Mpc. We however verified that,
even when taking the axis of the sample along the side of the box
and therefore accepting a small set of replicas,
the statistical results do not change. This is fully expected,
since at worst the points replicated (and then given double statistical
weight) are only those located in the tip of the ``wedge'' 
within a distance $l_{\rm
tip}\simeq d_{\rm lim}-l_{\rm box}=4\Mpch$ from the observer. 
This tiny region contains on average less
than 1 object ($l_{\rm tip} < d_{\rm gal}$) and, because of its wedge shape, 
allows very few sampling cells within its boundaries even for the
smallest sphere radii. 
For each case, we then construct 20 artificial volume--limited
samples with the same boundary shapes as the observational VLS and the same
number of objects (with a 2\% tolerance). 

In order to set the $M/L$ parameter of our galaxy identification scheme,
we require  the resulting
galaxy distribution to have a variance ${\bar\xi_2}(r)$ in agreement with
the one measured for the real sample. The variance is estimated by using the
same sampling cells as for the VPF described in Section 2 ($r$ is the radius
of the cell). In Figure$\, $2 we plot the ${\bar\xi_2}(r)$ 
for the PPS galaxies (error bars are 3--$\sigma$ bootstrap
errors over 20 resamplings) and that obtained for
the artificial ones by averaging over the 20--sample sets from the BSI
and CDM simulations for two different values of $M/L$: in both panels, 
the lower
(dotted) curves correspond to the $M/L$ values providing the best fits to the
observed data, which are 1200$\,h$ and 2400$\,h$ for BSI and CDM respectively
(we refer to these ``best--fitting'' galaxy populations as {\sl Gal}$_1$).
Curves corresponding to $M/L=900\,h$ (for BSI) and 1800$\,h$ (for CDM) are also
given ({\sl Gal}$_2$; dot--dashed lines), to show the sensitivity of the 
result to a 25\%
change of the assumed $M/L$ ratio. The mass, $M_{\rm l.h.}$,
of the lightest halo selected and 
the overdensity threshold $w_{\rm th}$ used in each case are reported in 
  Table$\,1$. The values of $w_{\rm th}$ are conspicuously 
smaller than those of Paper~A because of the different scale on which 
haloes are defined. 

The best--fit $M/L$ values will then be used in the following VPF analysis.
They are fairly high with respect to those suggested by observations of
galaxy groups (e.g., Ramella, Geller, \& Huchra 1989; Nolthenius 1993; Mamon
1993; Moore, Frenk, \& White 1993), although with quite large uncertainties.
However, a comparison with the $M/L$ found for CDM in Paper~A
confirms the strong dependence of $M/L$ on the size of the cell $l_{\rm c}$. 
As we discussed above, the absolute values of the
  best--fit  $M/L$ parameters we work out in the present analysis cannot be given physical
significance, but the ratios between 
values for different cosmological models 
can have a physical meaning.
The $M/L$ for BSI is about $0.4$ times  the one for CDM,
  approximately the same ratio that was                     
found between CHDM ($M/L\simeq 250\,h$)
and CDM ($M/L\simeq 600\,h$ for bias
  $b=1.0$ like here) in Paper A. Therefore, 
we expect BSI and CHDM models to have comparable $M/L$ values.
Since the simulations considered in Paper~A had $l_{\rm
  c}\simeq 0.1\Mpch$ and therefore provided more reliable estimates of $M/L$,
  this fact seems to favor the BSI model over    standard CDM 
  in reproducing observations, which point towards rather low values of $M/L$.
 However, the limited resolution
of our simulations does not allow us to draw a firm conclusion on this point.

To try to avoid the difficulties connected with the choice of $M/L$ and check
the robustness of our results, we will also perform our VPF analysis 
directly on the
halo population. In this case, each halo is regarded as a galaxy and assigned a
luminosity proportional to its mass. We take then the $N_{\rm hal,2}$ most
massive haloes such that $N_{\rm hal,2} = (l_{\rm box}/d_{\rm gal})^3$, where $d=5.5\Mpch$ is 
the average galaxy separation of real galaxies as above
(in the following we will refer to this halo
population as {\sl Hal}). The mass and overdensity 
thresholds we found are reported
in Table$\,1$, along with the $M/L$ ratios for our halo--based ``galaxies''.
Their variance $\bar\xi_2(r)$ is shown in Figure~3
(dashed line for BSI and dotted for CDM)
and is obtained as
usual by averaging over 20 observer's locations for each case. 
Both halo populations are less correlated than the PPS
galaxies (filled circles in the Figure), though
BSI data generally agree with observations within the
errors. CDM {\sl Hal} fares worse and displays a variance significantly
smaller than the real data on all scales below $6\Mpch$. 

Here we do not discuss the qualitative aspects of the
``galaxy'' distributions obtained from these BSI and CDM simulations.
A visual inspection of BSI vs. CDM has already been carried out by Kates et al.
(1995), by using a set of simulations which includes the present ones
(although their galaxy identification scheme was different).
Also, our artificial galaxy populations, being bound to mimic 
the bright galaxies of the PPS
sample, are rather sparse, thus making a visual approach quite poor. 
  However, our purpose here is to provide {\sl quantitative} estimates of
 the void distribution
in BSI and CDM in an {\sl observational context}, i.e. facing  the same
constraints as those holding for a real galaxy sample. This is what we attempt 
to achieve by the VPF analysis of the artificial VLSs.
 
\section{The VPF analysis}

In this Section, we evaluate the VPF for the set of real and artificial VLSs
according to the technique illustrated in Section$\,2$. 
Figure~4 shows the results for real (PPS) and simulated {\sl Gal}
galaxies, for BSI (panel $a$) and CDM (panel $b$). 
The VPF for real data is expected to be within the 
error bars plotted, 
which correspond to 3 times the bootstrap errors. 
VPFs obtained from our
Schechter--distributed {\sl Gal}$_1$ galaxies are shown
by 5 (dotted) curves, corresponding to 5 different observer's locations 
(selected at random among the 20--observer sets
considered for each model).
  Let us recall that {\sl Gal}$_1$ 
curves are obtained with the best--fit $M/L$ values of Figure$\,2$ (see
also Table$\,1$). For the sake of comparison, both in
panel $(a)$ and $(b)$, we also plot an observer--averaged  VPF curve
(dot--dashed line) obtained
for {\sl Gal}$_2$ galaxies, which have 
the alternative $M/L$ value considered in Figure $\,2$.
It is noticeable how
the choice of a value of $M/L$, performed just to obtain the observed 
${\bar\xi_2}(r)$
behavior, has a direct impact on the fit of the VPF with real data, namely
the $M/L$ which gives the best--fitting $\bar\xi_2$ also ensures 
a good fit to the observational VPF.
In the Figure, we also plot the Poissonian VPF as a long--dashed curve.

Both standard CDM and BSI models correctly reproduce the PPS data
within errors. 
This can be more carefully checked by examining 
Table$\,2$, where we report the VPF values at eight
 different scales along with the full {\sl sky  variance}  for the simulations,
 (i.e. the scatter 
  among the VPFs  measured by the 20 different observers).

Figure~5 allows us a test of the dependence of our results on the
  galaxy identification scheme. The Figure shows the dependence
  of the VPF on   $r$ for the halo populations ({\sl Hal}) defined in the
previous Section, both for BSI and CDM.
The error bars are still
3--$\sigma$ bootstrap errors for the real sample, whereas the short--dashed 
and dotted curves refer to the VPFs averaged over 20
observers for BSI and CDM respectively. 
As a reference, we always plot the Poissonian VPF (long--dashed). 
Clearly, the BSI model provides a good fit to the observational data also when
haloes are directly treated as galaxies, whereas the {\sl Hal} curve for
  CDM tends to diverge slightly from
the PPS one at scales below $3\Mpch$. Nonetheless, in the intermediate scale
range ($3\Mpch <r<6\Mpch$), where the VPF discriminates
between CHDM and PPS as shown in Paper~A, the CDM curve is still in 
agreement with observational data, and BSI data as well. The fact that no
difference between BSI and CDM is found in that range of scales is
an important point. In fact, on these scales, because of the  
coupling among density fluctuations of different wavelength in the
non--linear regime (see Kauffmann \& Melott 1992, in particular), 
we could expect the feature in the BSI spectrum (occurring at $r\simeq
\lambda_{\rm break} = 9.4\,h^{-1}$Mpc) to start playing its effects and cause
deviations between the VPFs of the two models.

\section{Discussion}
 
In this paper we debated
the void probability function statistics as a discriminator between models
on the scales of non--linear clustering ($1\Mpch\le r\,\mincir 8\Mpch$).
We considered $N$--body simulations of BSI and standard CDM 
  (both normalized to first--year COBE data) and estimated the
redshift--space VPF for artificial galaxy samples so as to perform 
a close comparison with a large volume--limited sample of the
Perseus--Pisces survey. Artificial galaxies 
were suitably identified as residing inside peaks 
of the {\sl evolved} density field. 
We tested the robustness of our results against the details of the
  galaxy identification method, by also considering the simplest scheme 
in which each peak is directly associated with a galaxy.
Of course, this does not compensate for our ignorance on how galaxies actually
form and our results should still be regarded with some caution.

We showed that both BSI and standard CDM fit
observational data quite well. For the CDM, this confirms 
the findings of Ghigna et al. (1994, Paper~A). As mentioned in the
previous Section, the fact that the VPF
results for BSI can match those for CDM so well is rather remarkable in
view of the feature exhibited by the linear spectrum of density
fluctuations in the BSI model. 
As is known, in the
non--linear evolution of density fluctuations components characterized by
different wavelengths have no longer a separate evolution, and a feature on
$\lambda_{\rm break}\simeq 9.4\Mpch$ should have an influence on
 a fairly wide range of scales, at least down to $\sim \lambda_{\rm break}/3$ 
(Kauffmann and Melott 1992). In our VPF analysis this spectral feature has 
no significant effect. 

In Paper~A it was also found that the VPF can discriminate between CDM and 
a CHDM model with $\Omega_{\rm baryon}/\Omega_{\rm cold}/\Omega_{\rm hot} =
0.1/0.6/0.3$ and one neutrino flavor. This, together with the fact that both
unbiased and biased ($b=1.5$) CDM had the same VPF, was taken as an
indication that the void statistics is directly and mostly 
sensitive to the {\sl shape}
of the linear spectrum and/or the nature of the dark matter.
Here, we could separately test
its dependence on the  shape of the spectrum, since 
BSI and CDM are both based on pure cold dark
matter. Given that our results do not show such a dependence, 
we can tentatively propose the following conjecture: The statistics of
voids in the non--linear clustering regime 
can be influenced by the nature (i.e. the composition) 
of DM more effectively than by the shape
of the density fluctuation spectrum. A dominant role in determining the
void distribution could be played by the
velocity dispersion of hot particles and by their ability (or difficulty) to
follow cold dark matter when it clusters. 

However, there seems to 
exist a complicated dynamical interplay. A counter--example is 
provided by a CHDM model with
$\Omega_{\rm baryon}/\Omega_{\rm cold}/\Omega_{\rm hot} = 0.05/0.75/0.2$ 
and two equal--mass neutrino flavors
(Primack et al. 1995). In that
case, which involves even more neutrinos carrying a
non--vanishing mass, 
preliminary results on the VPF indicate a good agreement with PPS data (Ghigna 
et al. 1996). This case also represents an instance of
how our conjecture can be used to constrain the ``particle physics'' at
the basis of a given cosmological model. In fact, if our conjecture is 
true, the agreement shown by the two--neutrino model
could be a point specifically in support of such a ``recipe'' for the
composition of mixed dark matter. 
 Anyway, as far as the analysis performed in this paper is concerned,
there is clearly no way to say which one is better among the models
passing the VPF test. 

\acknowledgments

We wish to thank the referee for his careful reviewing of the paper which
greatly helped us to improve its presentation.
Thanks are also due to Stefano Borgani for a number of discussions on specific
points. S.~Ghigna is grateful to Durham University for giving him access
to the Starlink facility and acknowledges use of its editing software
 during the completion
of the paper. J.~R.~thanks the Physics Department of Milano University for its 
hospitality during the preparation of this work and acknowledges the financial 
support of E.~C. through the HCM network program.

\clearpage   

\begin{table*}
\begin{center}
\begin{tabular}{lrrcrrrrr}
\tableline       
Model~ & Scheme~~ & $M/L$  &  $M_{\rm l.h.}/M_\odot$ & $w_{\rm th}$ &
$N_{\rm hal}$\\
\tableline       
BSI:~& {\sl Gal}$_1$~~ &   $1200\,h$ &  $4.2\times 10^{12}\,h^{-1}$ & 86 & 2115\\
\null  & {\sl Gal}$_2$~~ &   $900\,h$ & $6.7\times 10^{12}\,h^{-1}$ & 130& 1112 \\
\null & {\sl Hal}~~~ &   $1400\,h$ & $3.7\times 10^{12}\,h^{-1}$& 77& 2534\\
\tableline       
CDM:~ & {\sl Gal}$_1$~~ &   $2400\,h$ & $8.5\times 10^{12}\,h^{-1}$ & 110 & 1865\\
\null & {\sl Gal}$_2$~~ &   $1800\,h$ & $1.5\times 10^{13}\,h^{-1}$ &290& 664 \\
\null & {\sl Hal}~~~ &   $2800\,h$ & $3.6\times 10^{12}\,h^{-1}$ & 80 & 2534\\
\tableline       
\end{tabular}   
\end{center}

\caption{Values of the $M/L$ parameter, $M_{\rm l.h.}$ (mass of the lightest 
  halo associated with a galaxy), and $w_{\rm th}$ (overdensity threshold)
 for different galaxy identification schemes,
for BSI and CDM. {\sl Gal} refers to Schechter--distributed 
galaxies obtained from DM haloes through fragmentation, after
 assuming a value for $M/L$: {\sl Gal}$_1$ galaxies provide the best
fit to the observational ${\bar\xi_2}$;
{\sl Gal}$_2$ ones are obtained after a 25\% increase of $M/L$. 
{\sl Hal} refers to DM haloes directly associated with galaxies.
  $N_{\rm hal}$ is the number of haloes used              in
  the simulation volume for each case. The connection of $M/L$ with the physical
mass--to--light ratio is discussed in Section~5.}
\end{table*}

\begin{table*}
\begin{center}
\begin{tabular}{crrrrrr}
\tableline       
  $r$ ($h^{-1}$Mpc)  & \multicolumn{5}{c}{$P_0(r)\times 10^2$}\\
\tableline       
 & {PPS~~~~} & {BSI: {\sl Gal}$_1$} & {BSI: {\sl Hal}~} 
   & {CDM: {\sl Gal}$_1$} & {CDM: {\sl Hal}}\\
$2.04$  & $86.3\pm0.7$     & $86.3\pm2.0$ & $85.8\pm1.9 $ 
  & $86.7\pm2.5$  & $84.7 \pm 1.5$  \\
$2.42$  & $79.9\pm1.2$     & $80.4\pm3.7$ & $78.9\pm 2.7$
  & $80.6 \pm3.1 $  & $ 77.9 \pm 3.4 $  \\
$2.85$  & $70.1\pm 1.2$    & $72.1\pm5.4$  & $71.3\pm3.8 $ 
  & $72.5 \pm5.5$  & $ 69.4 \pm 4.4$    \\
$3.37$  & $59.3\pm 3.0$    & $62.8\pm6.7$  & $60.9\pm 7.4$
  & $61.5 \pm5.4$  & $ 58.8 \pm 6.3$ \\
$3.98$  & $49.4\pm3.5 $    & $52.0\pm8.8$ & $49.7\pm 9.8$
  & $49.6\pm5.4$  & $47.1 \pm 8.9$ \\
$4.70$  & $36.2\pm3.2$    & $37.5\pm7.2$  & $36.6\pm9.5 $ 
  & $36.2\pm8.5$ & $ 35.2\pm 10.6$ \\
$5.56$  & $25.7\pm 4.2$    & $25.4\pm7.9$ & $23.8\pm 9.6$ 
  & $ 23.5\pm 7.3$ & $ 23 \pm 12 $ \\
$6.57$  & $11.5\pm6.5$     & $15\pm12$  & $13 \pm 10$ 
  & $14 \pm 10$ & $ 13 \pm 16$ \\
\tableline       
\end{tabular}   
\end{center}

\caption{The VPF, $P_0$, at various scales $r$ for observational data
(PPS), for ``best--fitting'' artificial galaxies ({\sl Gal}$_1$) and DM
haloes ({\sl Hal}) in the simulations.
Errors are 3 standard deviations over 20 bootstrap 
  resamplings  for the observational data, and over 20
  different observers  for simulated samples ({\sl sky  variance}).}
\end{table*}

\clearpage


\begin{figure}
\caption{Linear power spectrum of density fluctuations for the 
standard CDM model 
and the BSI model (with the parameters considered in the text), both
normalized to first--year COBE data.  The double bar near the bottom of
the plot indicates the dynamical range covered by the
simulations used in this paper.
The size of the computational box, $L=75 \Mpch$, yields 
$k_{\rm{min}}=2\pi/L$, while the Nyquist wavenumber of the grid gives
$k_{\rm{max}}=k_{\rm{min}}N_{\rm{g}}/2\,$ ($N_{\rm{g}}=256$).}
\end{figure}

\begin{figure}
\caption{Variance ${\bar\xi_2}$ vs. scale $r$ (radius of the spherical 
sampling volumes) for real (PPS; error bars are 3--$\sigma$ bootstrap
errors) and simulated galaxies ({\sl Gal}$_1$ and {\sl Gal}$_2$). Panels $(a)$ and $(b)$ are for BSI
and CDM models, respectively. {\sl Gal} galaxies are described in Table$\,1$.}
\end{figure}

\begin{figure}
\caption{Variance $\bar\xi_2$ vs. $r$ for DM haloes ({\sl Hal}) compared
with the PPS one, both for BSI and CDM.}
\end{figure}

\begin{figure}
\caption{Void probability function $P_0$ vs. $r$ for observational
data (PPS), for five artificial VLSs of {\sl Gal}$_1$ galaxies and
averaged over the 20 VLSs of {\sl Gal}$_2$ galaxies. Panel~$(a)$ is for 
BSI and $(b)$ is for CDM. The Poissonian curve refers to a completely
uncorrelated distribution.}
\end{figure}

\begin{figure}
\caption{VPF vs. $r$ for DM haloes ({\sl Hal}) compared with the PPS one,
both for BSI and CDM.}
\end{figure}

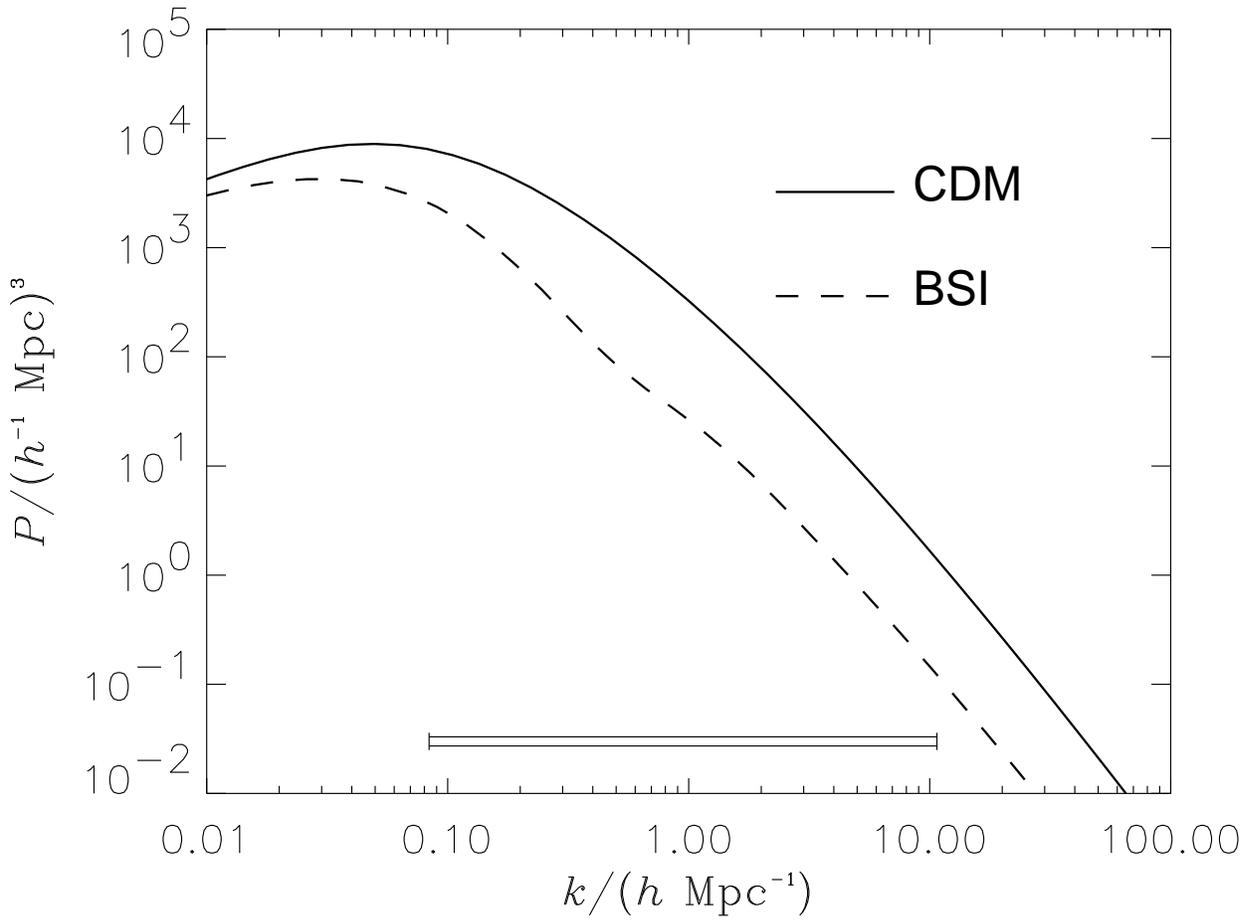
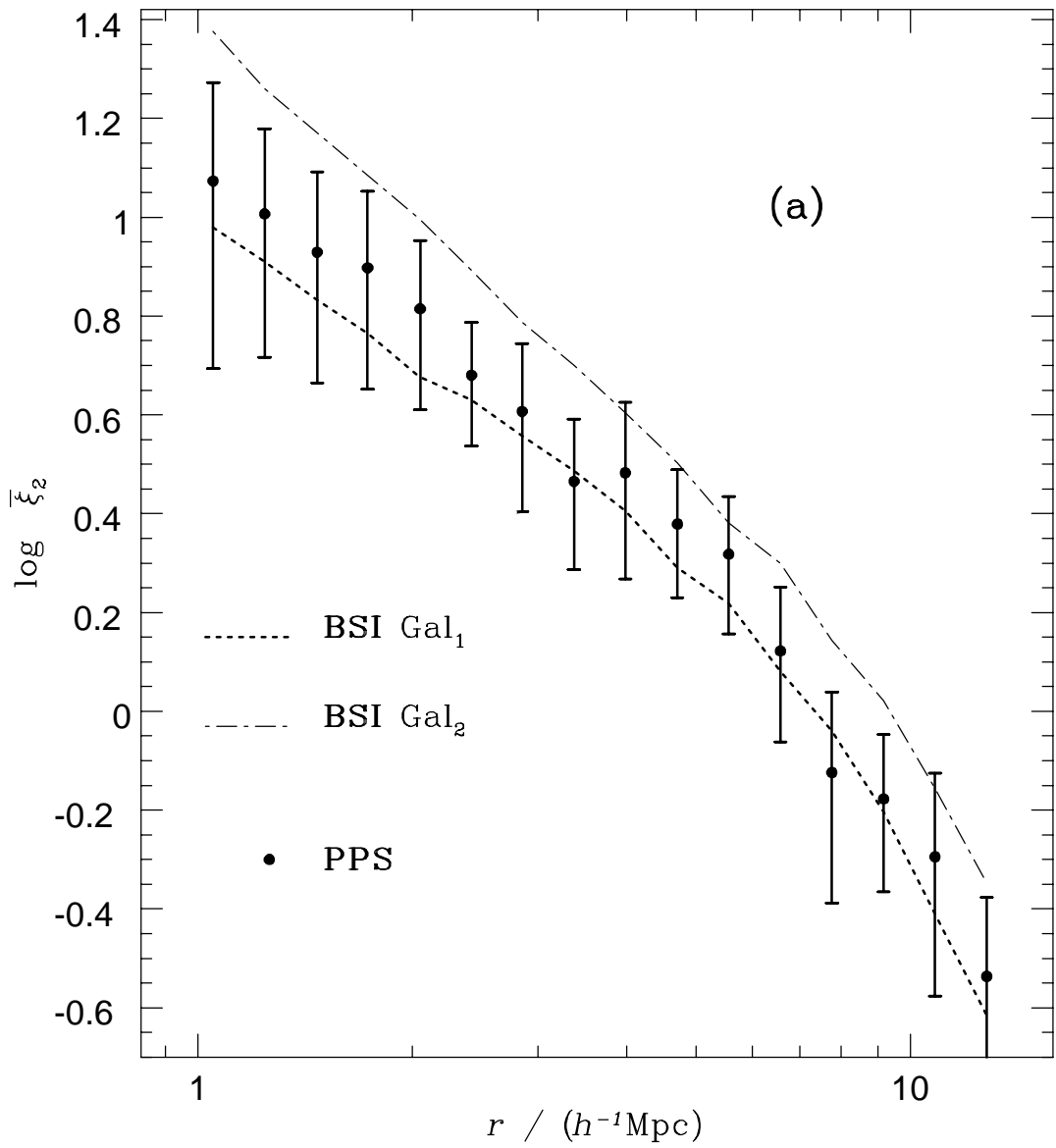
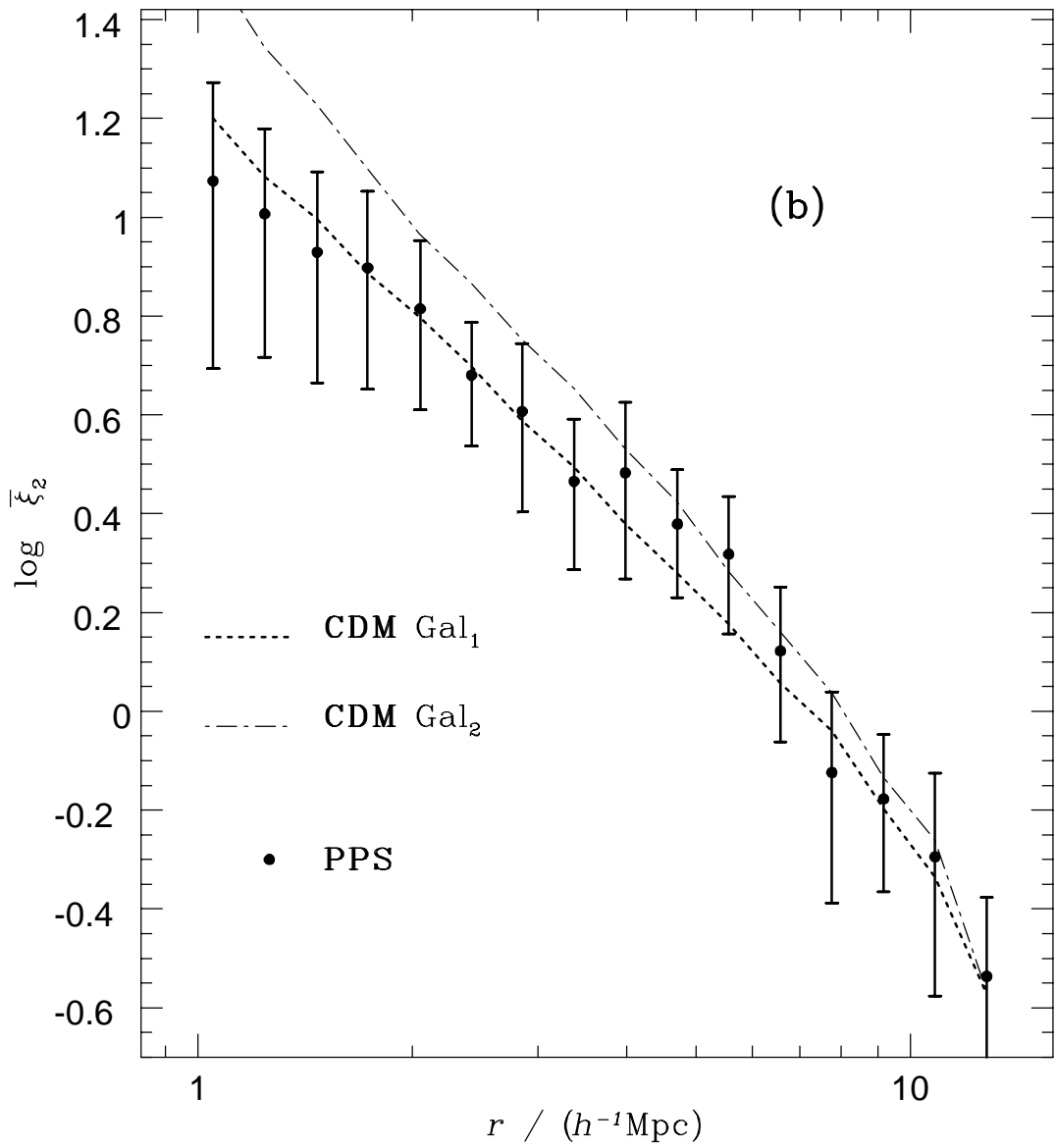
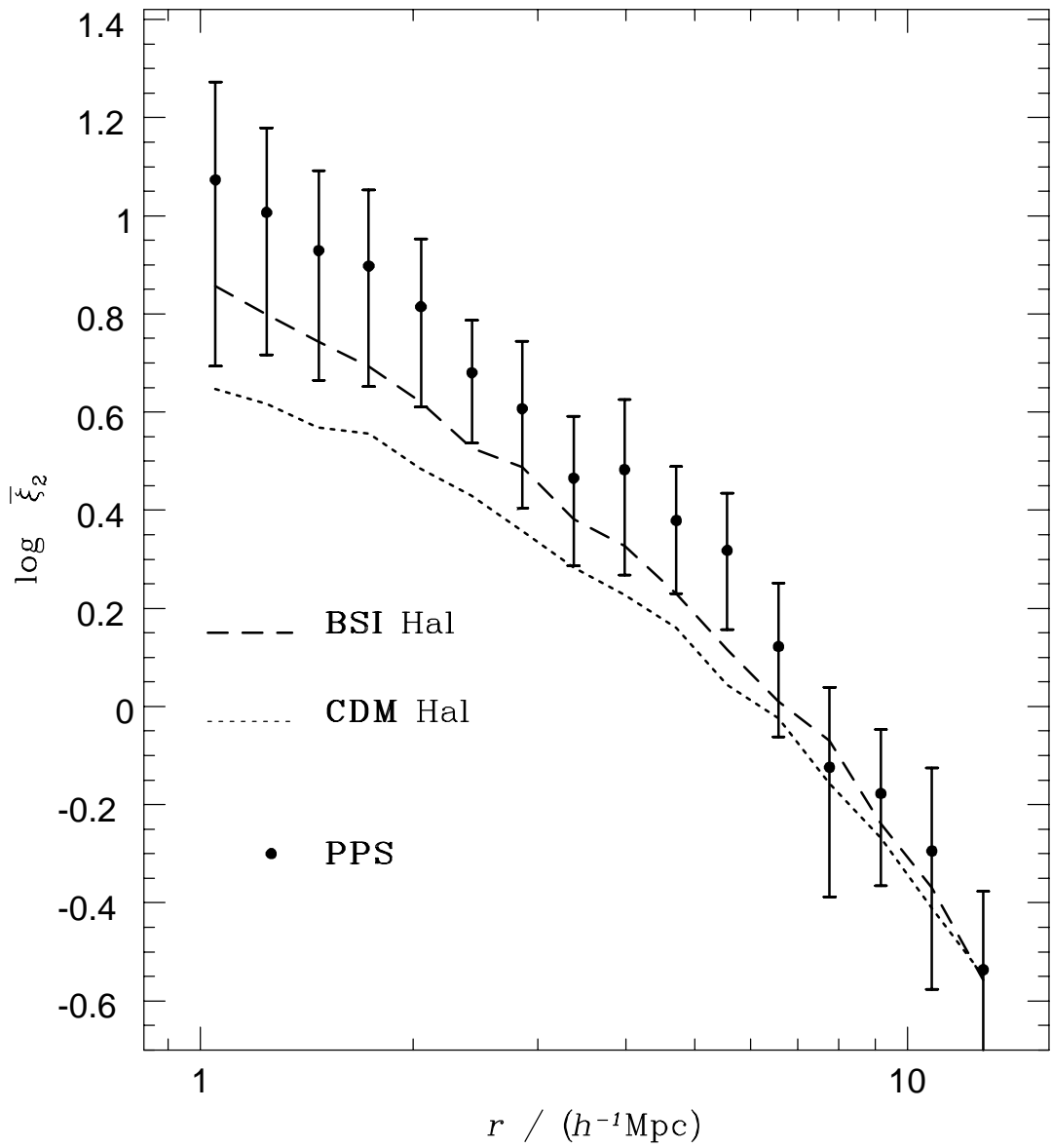
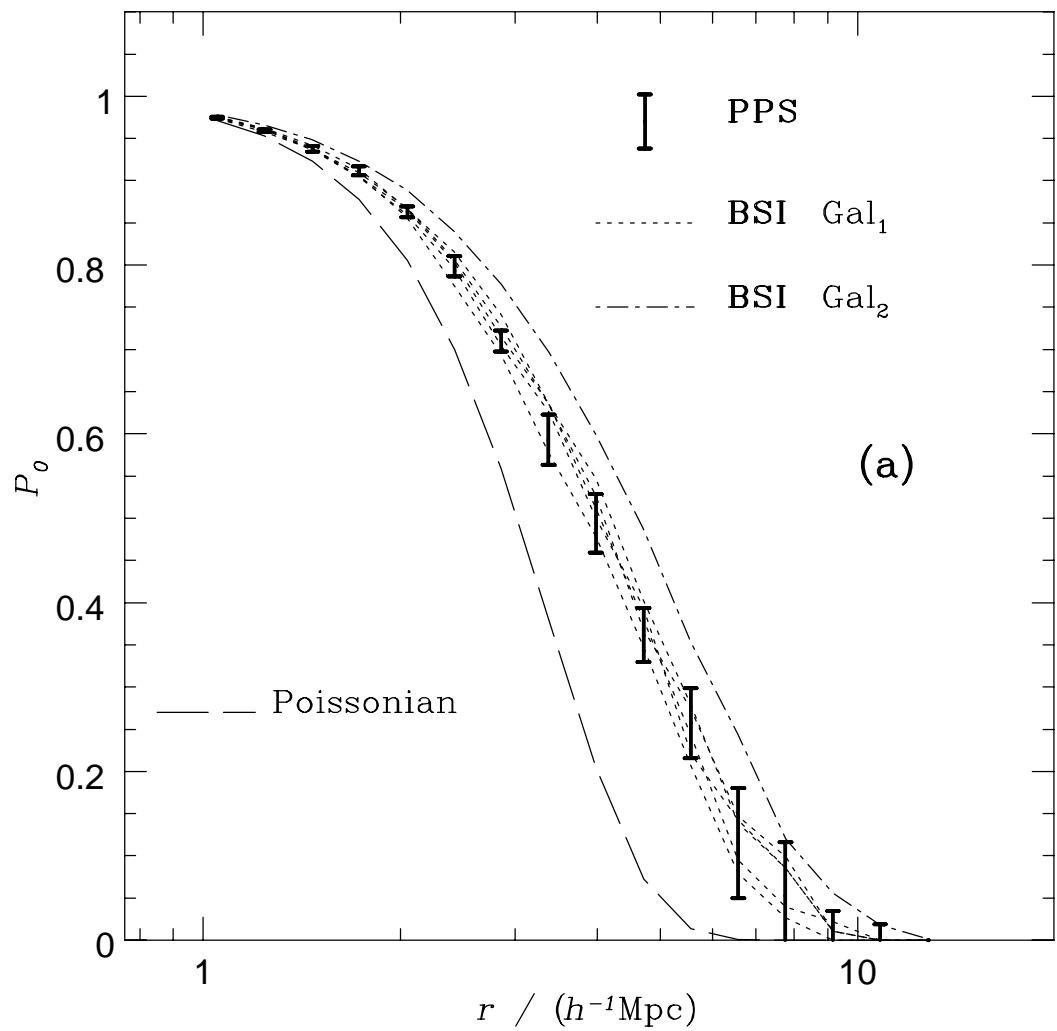
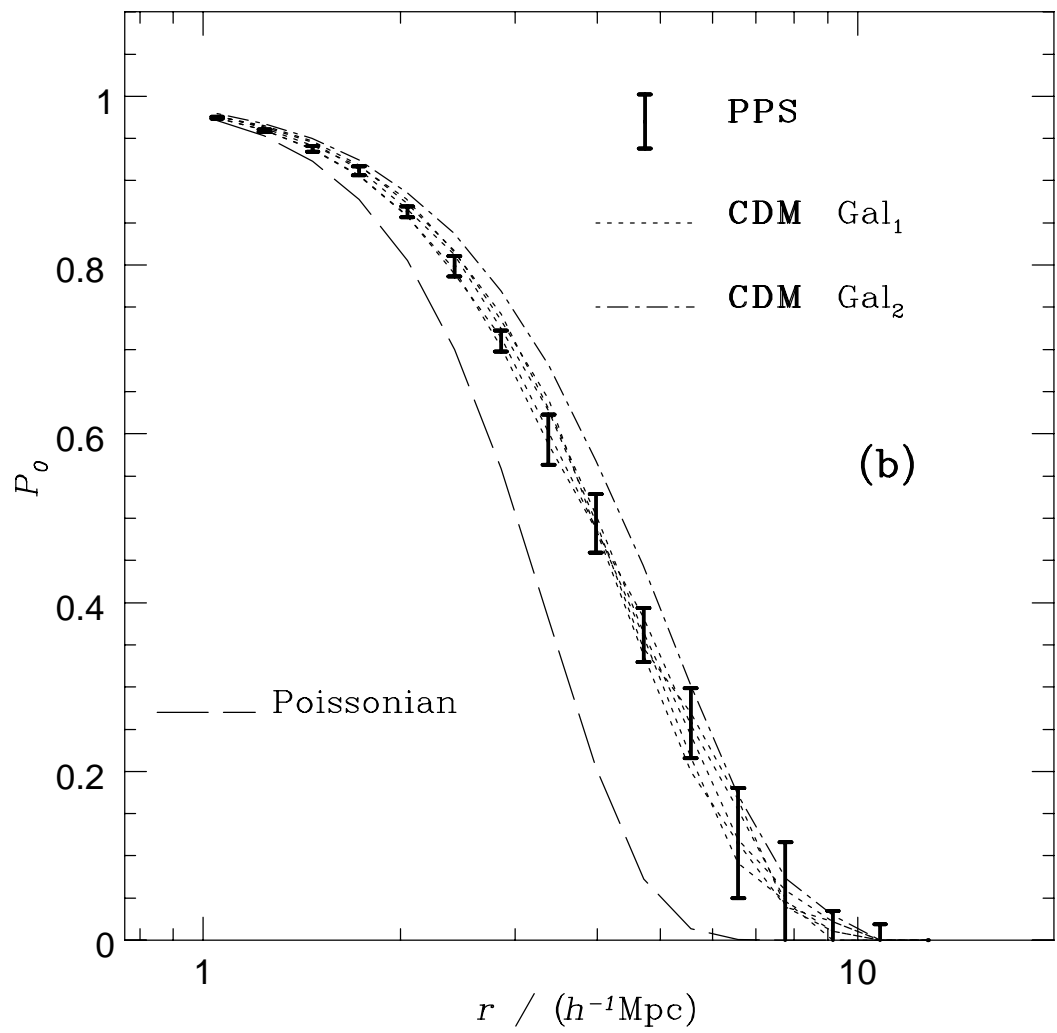
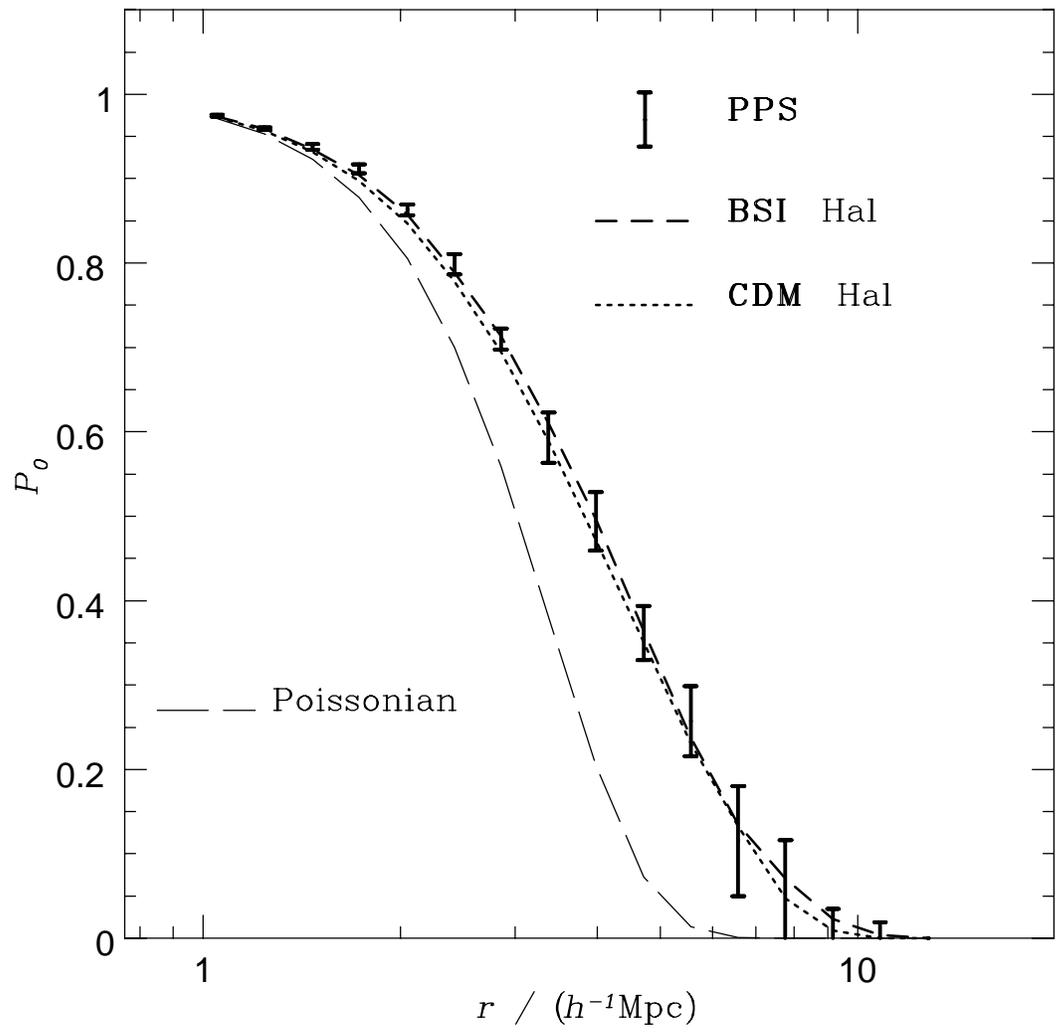
\end{document}